\documentclass[aps,floats,floatfix,reprint,superscriptaddress,twocolumn,nobalancelastpage,final]{revtex4-2}

\usepackage[T1]{fontenc}
\usepackage[utf8]{inputenc}
\usepackage{graphicx}
\usepackage{dcolumn,bm,float}
\usepackage{amssymb,amsmath,amsfonts}
\usepackage{color,xcolor,url}
\usepackage{ragged2e}
\usepackage{booktabs}
\usepackage{lipsum}
\usepackage[percent]{overpic}
\usepackage{amsmath}
\usepackage{amsfonts}
\usepackage{mathtools}

\usepackage{hyperref}
\hypersetup{colorlinks=false,
    linkcolor=black,
    citecolor=black,
    urlcolor=black,
    pdftitle={Flow Divergence: Comparing Maps of Flows with Relative Entropy},
    pdfauthor={Christopher Bl{\"o}cker, Ingo Scholtes}
}
\usepackage{xurl}
\usepackage[capitalise]{cleveref}

\usepackage{orcidlink}

\setcitestyle{numbers,comma,sort&compress,open={[},close={]}}

\begin{document}

\preprint{APS/123-QED}

\title{Flow Divergence: Comparing Maps of Flows with Relative Entropy}

\author{Christopher Bl{\"o}cker~\orcidlink{0000-0001-7881-2496}}
\affiliation{%
 \mbox{Data Analytics Group, Department of Informatics, University of Zurich, CH-8006 Zurich, Switzerland}
}%
\affiliation{%
 \mbox{Chair of Machine Learning for Complex Networks, Center for Artificial Intelligence and Data Science},
 \mbox{University of W{\"u}rzburg, DE-97070 W{\"u}rzburg, Germany}
}%

\author{Ingo Scholtes~\orcidlink{0000-0003-2253-0216}}
\affiliation{%
 \mbox{Chair of Machine Learning for Complex Networks, Center for Artificial Intelligence and Data Science},
 \mbox{University of W{\"u}rzburg, DE-97070 W{\"u}rzburg, Germany}
}%
\affiliation{%
 \mbox{Data Analytics Group, Department of Informatics, University of Zurich, CH-8006 Zurich, Switzerland}
}%

\date{\today}%

\begin{abstract}
Networks represent how the entities of a system are connected and can be partitioned differently, prompting ways to compare partitions.
Common approaches for comparing network partitions include information-theoretic measures based on mutual information and set-theoretic measures such as the Jaccard index.
These measures are often based on computing the agreement in terms of overlap between different partitions of the same set.
However, they ignore link patterns which are essential for the organisation of networks.
We propose flow divergence, an information-theoretic divergence measure for comparing network partitions, inspired by the ideas behind the Kullback-Leibler divergence and the map equation for community detection.
Similar to the Kullback-Leibler divergence, flow divergence adopts a coding perspective and compares two network partitions $\mathsf{M}_a$ and $\mathsf{M}_b$ by considering the expected extra number of bits required to describe a random walk on a network using $\mathsf{M}_b$ relative to reference partition $\mathsf{M}_a$.
Because flow divergence is based on random walks, it can be used to compare partitions with arbitrary and different depths.
We show that flow divergence distinguishes between partitions that traditional measures consider to be equally good when compared to a reference partition.
Applied to real networks, we use flow divergence to estimate the cost of overfitting in incomplete networks and to visualise the solution landscape of network partitions.
\end{abstract}

\maketitle

\section{Introduction}
Many real-world complex networks have communities: groups of nodes that are more linked to each other than to the rest.
Communities capture link patterns and abstract from groups of individual nodes, revealing how networks are organised at the mesoscale.
For example, tightly-knit groups of friends in social networks, groups of interacting proteins in biological networks, or traders who perform transactions in financial networks form communities.
Motivated by various use cases and based on different assumptions, researchers have proposed a plethora of ways to characterise what constitutes a community, however, none of these characterisations is fundamentally more right or wrong than any other~\cite{FORTUNATO201075}.
Naturally, based on their assumptions, different methods partition the same network into communities differently; and running a stochastic method on the same network several times can return different partitions.
Consequently, we need measures to compare partitions and evaluate to what extent they agree and how they differ.

Researchers across scientific fields have proposed partition similarity measures~\cite{vanderHoef2019,jaccard-index,cover2006elements,Straulino2021,10.1371/journal.pone.0292018,pmlr-v32-romano14,PhysRevE.92.062825,donnat2018tracking}.
Arguably, the most commonly used measures in network science are the so-called Jaccard index~\cite{jaccard-index} and information-theoretic measures based on mutual information~\cite{cover2006elements,pmlr-v32-romano14,PhysRevE.92.062825}.
These measures have been applied successfully to a wide range of research questions, including applications in economics~\cite{Straulino2021}, bioinformatics~\cite{Holmgren2023}, and social network analysis~\cite{doi:10.1137/080734315}.
However, in the context of community detection, they have a crucial shortcoming: while community detection in complex networks is about grouping together nodes with similar link patterns, popular partition similarity measures ignore links altogether.
Instead, they merely consider how well the communities from different partitions coincide, essentially measured in terms of set overlaps~\cite{esquivel2012comparing}.

We address this shortcoming of existing partition similarity scores by developing a measure based on the description of random walks on networks to incorporate link patterns.
Combining the principles behind the Kullback-Leibler (KL) divergence~\cite{cover2006elements} for measuring the dissimilarity between stochastic processes and the map equation for community detection~\cite{doi:10.1073/pnas.0706851105}, our measure quantifies the expected extra number of bits required to describe a random walk when using a network partition $\mathsf{M}_b$ relative to a reference partition $\mathsf{M}_a$.
Flow divergence relates to map equation similarity \cite{pmlr-v198-blocker22a}, an information-theoretic node-similarity measure that quantifies the required number of bits for describing a random-walker step, given a network partition.
Because flow divergence is based on describing random walks, our measure naturally takes the network's link patterns into account and can compare two-level and hierarchical partitions with each other.
Moreover, we can identify on a per-node basis how much each node's community assignment contributes to the divergence, allowing us to distinguish between high and low-impact assignments.

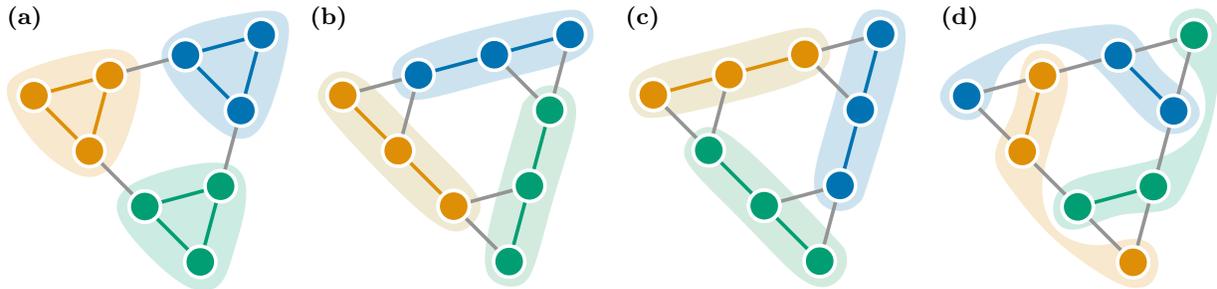
\begin{figure*}[ht!]
    \centering
    \begin{overpic}[width=.9\linewidth]{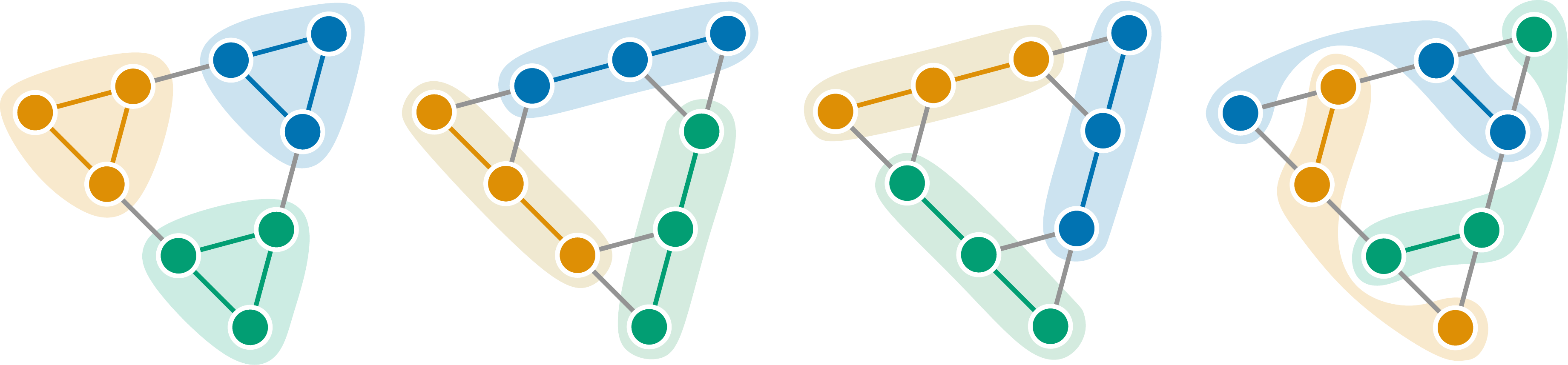}
        \put(0,22){\textbf{(a)}}
        \put(25,22){\textbf{(b)}}
        \put(51,22){\textbf{(c)}}
        \put(77,22){\textbf{(d)}}
    \end{overpic}
    \caption{Four different partitions for the same network.
    Because common measures such as the Jaccard index and mutual information, including its variants, consider merely node labels but ignore link patterns, they consider partitions \textbf{(b)}, \textbf{(c)}, and \textbf{(d)} as equally good when compared against the reference partition \textbf{(a)}.}
    \label{fig:shortcomings}
\end{figure*}
\section{Background}
Comparing partitions is a recurring problem in applications across scientific domains and has received much attention:
During the last one hundred and fifty years or so, researchers have proposed a plethora of measures for comparing partitions~\cite{vanderHoef2019,jaccard-index,cover2006elements,Straulino2021,10.1371/journal.pone.0292018,pmlr-v32-romano14,PhysRevE.92.062825,donnat2018tracking}.
Given two partitions $\mathcal{A}$ and $\mathcal{B}$ of the same set $\mathcal{X}$ with $n = \left|\mathcal{X}\right|$ objects, the aim is to measure how similar those two partitions are.
In the context of method development, one of the two partitions, say $\mathcal{A}$, is often a known ground truth partition against which the new method's prediction, $\mathcal{B}$, is evaluated.

Two of the arguably most commonly used partition similarity measures in network science are the Jaccard index~\cite{jaccard-index} and information-theoretic scores based on mutual information~\cite{cover2006elements}.
The Jaccard index computes the agreement between groups $A \in \mathcal{A}$ and $B \in \mathcal{B}$ as $J\left(A,B\right) = \frac{\left|A \cap B\right|}{\left|A \cup B\right|}$.
Computing the similarity between partitions $\mathcal{A}$ and $\mathcal{B}$ requires finding the best match $B \in \mathcal{B}$ for each $A \in \mathcal{A}$ and weighing according to $A$'s size: $J\left(\mathcal{A},\mathcal{B}\right) = \sum_{A \in \mathcal{A}} \frac{\left|A\right|}{n} \max_{B \in \mathcal{B}} J\left(A,B\right)$.
Mutual information, which is the basis for many information-theoretic measures, considers how much uncertainty remains about the assignments of objects under $\mathcal{B}$, given that their assignment under $\mathcal{A}$ is known, $I\left(\mathcal{A}, \mathcal{B}\right) = \sum_{A \in \mathcal{A}} \sum_{B \in \mathcal{B}} P\left(A,B\right) \log_2 \frac{P\left(A, B\right)}{P\left(A\right)P\left(B\right)}$, where $P\left(A\right) = \frac{\left|A\right|}{n}$ and $P\left(B\right) = \frac{\left|B\right|}{n}$ are the probabilities of selecting $A$ and $B$, respectively, when choosing an object $X \in \mathcal{X}$ uniformly at random; $P\left(A,B\right) = \frac{\left|A \cap B\right|}{n}$ is the joint probability of $A$ and $B$.
Normalised mutual information scales the mutual information score to the interval $\left[0,1\right]$, and adjusted mutual information (AMI) additionally adjusts it for chance \cite{pmlr-v32-romano14}.

The Jaccard index and mutual information ignore link patterns because they only consider group memberships, which, depending on the application, may be a valid approach.
However, to see why this is a problem when comparing communities that summarise link patterns, consider the example network shown in \cref{fig:shortcomings}: a network with nine nodes, partitioned in four different ways.
We assume that partition $\mathcal{A}$ (\cref{fig:shortcomings}a) represents the network's true community structure and compare partitions $\mathcal{B}$, $\mathcal{C}$, and $\mathcal{D}$ (\cref{fig:shortcomings}b-d) against $\mathcal{A}$.
The Jaccard index is oblivious to the alternative partitions and judges them to match the reference partition equally well, $J\left(\mathcal{A}, \mathcal{B}\right) = J\left(\mathcal{A}, \mathcal{C}\right) = J\left(\mathcal{A}, \mathcal{D}\right) = \frac{1}{2}$.
Mutual information suffers from the same issue, $I\left(\mathcal{A}, \mathcal{B}\right) = I\left(\mathcal{A}, \mathcal{C}\right) = I\left(\mathcal{A}, \mathcal{D}\right) \approx 0.46$.
However, when evaluated with community detection in mind, it seems plausible that $\mathcal{B}$ and $\mathcal{C}$ agree with $\mathcal{A}$ to the same extent because of symmetry.
But partition $\mathcal{D}$ should be distinguished from $\mathcal{B}$ and $\mathcal{C}$: it has a different shape and captures a different pattern.

Despite ignoring links, measures such as the Jaccard index and mutual information are often used to compare network partitions whose groups summarise link patterns according to some community-detection objective function such as modularity \cite{doi:10.1073/pnas.0601602103} or the map equation \cite{doi:10.1073/pnas.0706851105}.
However, most commonly, community detection is about identifying groups of nodes with high within-group link density and low between-group density.
That is, link patterns play an important role in deciding what partition constitutes a good description of a network and should, therefore, be considered when comparing partitions.
\citet{Straulino2021} recognised this issue and proposed a framework to compare network partitions based on any community-detection objective function $f$ that takes a graph and a partition of its nodes, and returns, possibly after scaling, a value in the range $\left[0,1\right]$.
Treating $f$ as a black box, they define the non-symmetric distance between two partitions $\mathcal{A}$ and $\mathcal{B}$ as $d\left(\mathcal{A}, \mathcal{B}\right) = 1 - \frac{f\left(G, \mathcal{A}\right)}{f\left(G, \mathcal{B}\right)}$, where $G$ is a graph, thereby incorporating link patterns into their score.

\begin{figure*}[ht!]
    \centering
    \begin{overpic}[width=1\linewidth]{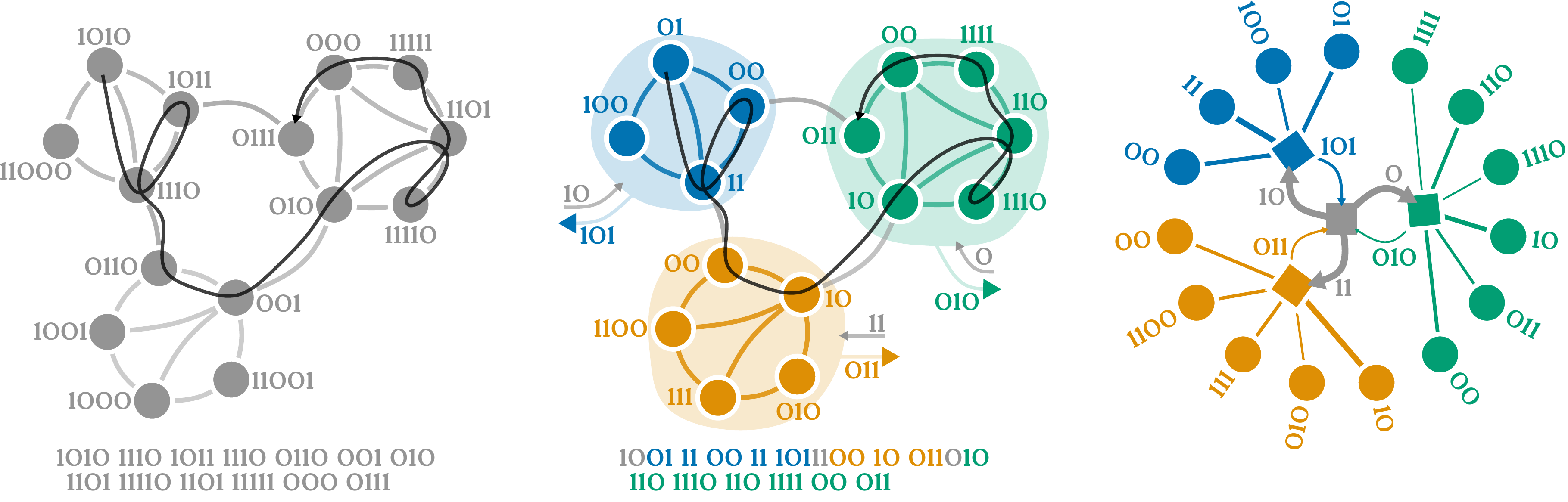}
        \put(0,32){\textbf{(a)}}
        \put(35,32){\textbf{(b)}}
        \put(71,32){\textbf{(c)}}
    \end{overpic}
    \caption{Illustration of encoding random walks and the principles behind the map equation.
    \textbf{(a)} Nodes are not partitioned into communities. We derive unique codewords from the nodes' visit rates and use them to describe the shown random-walk sequence with the codewords at the bottom.
    \textbf{(b)} Nodes are partitioned into three communities and receive codewords that are unique within each community. Codewords for entering and exiting communities are shown next to arrows that point into and out of the communities.
    \textbf{(c)} The map corresponding to the community structure and coding scheme from \textbf{(b)} drawn as a radial tree. Link widths are proportional to module-normalised codeword usage rates. Good maps have small module exit rates.
    }
    \label{fig:map-equation-principle}
\end{figure*}

\section{Flow Divergence}
To define our partition dissimilarity score, flow divergence, we combine the map equation for flow-based community detection with the Kullback-Leibler (KL) divergence \cite{cover2006elements}, which is defined as
\begin{equation}
    D_{KL}\left(P \,||\, Q\right) = \sum_{x \in X} p_x \log_2 \frac{p_x}{q_x}.
    \label{eq:kl-divergence}
\end{equation}
Here $P$ and $Q$ are probability distributions that are defined on the same sample space $X$.
The KL-divergence quantifies the expected additional number of bits required to describe samples from $X$ using an estimate $Q$ of its true frequencies $P$.
Following this idea, we define flow divergence to quantify the expected additional number of bits required to describe a random walk on a network using an estimate $\mathsf{M}_b$ of its true partition $\mathsf{M}_a$.

\subsection{Random-Walk Description Length}
Let $G = \left(V, E, \delta\right)$ be a connected graph with nodes $V$, links $E$, and link weights $\delta\colon E \to \mathbb{R}^+$.
Further, let $P = \left\{ p_v \,|\, v \in V \right\}$ be the set of ergodic node visit rates.
According to Shannon's source coding theorem~\cite{shannon1948mathematical}, describing the random walker's position on the graph requires at least,
\begin{equation}
    H\left(P\right) = -\sum_{v \in V} p_v \log_2 p_v~\text{bits},
    \label{eq:entropy}
\end{equation}
where, $H$ is the Shannon entropy and $\log_2 p_v$ is the length of node $v$'s codeword in bits.
We can design concrete codewords with a Huffman code~\cite{huffman1952method} as shown in \Cref{fig:map-equation-principle}a, however, to develop flow divergence, we only require the codewords' theoretical lengths.
If the graph is strongly connected, we calculate the nodes' visit rates using the power iteration method to solve the recursive set of equations
\begin{equation}
    p_v = \sum_{u \in V} p_u t_{uv},
    \label{eq:visit-rates}
\end{equation}
where $t_{uv} = \frac{\delta\left(u,v\right)}{\sum_{v \in V} \delta\left(u,v\right)}$ is the probability that a random walker at node $u$ steps to node $v$; in weakly connected graphs, we can use the PageRank algorithm~\cite{doi:10.1137/140976649} or so-called smart teleportation~\cite{PhysRevE.85.056107} to calculate the visit rates.

By combining \Cref{eq:visit-rates,eq:entropy} and reordering, we explicitly relate describing the random walker's position to transitions along links,
\begin{equation}
    H\left(P\right) = -\sum_{u \in V} p_u \sum_{v \in V} t_{uv} \log_2 p_v.
    \label{eq:as-random-walk}
\end{equation}
However, this formulation adopts a global perspective with globally unique codewords for nodes without taking the network's community structure into account.
To turn the coding dependent on communities, also called modules, we introduce a parameter $\mathsf{M}$ that represents a partition of the network's nodes into modules.
We denote the module-dependent transition probability for stepping from $u$ to $v$ as $s\left(\mathsf{M}, u, v\right)$ such that $\log_2 s\left(\mathsf{M}, u, v\right)$ is the cost in bits for encoding a step from $u$ to $v$,
\begin{equation}
    H\left(\mathsf{M}\right) = -\sum_{u \in V} p_u \sum_{v \in V} t_{uv} \log_2 s\left(\mathsf{M}, u, v\right).
    \label{eq:as-modular-random-walk}
\end{equation}
Setting $s\left(\mathsf{M}, u, v\right) = p_v$ maintains the global coding perspective while setting $s\left(\mathsf{M}, u, v\right) = t_{uv}$ adopts a node-local coding perspective.
Adopting an intermediate modular perspective means taking $u$'s and $v$'s modular context into account when computing coding costs.

\subsection{The Map Equation for Modular Coding}
The map equation~\cite{doi:10.1073/pnas.0706851105,map-equation-review} is an information-theoretic objective function for flow-based community detection that provides a way to define a modular coding scheme.
The map equation identifies communities by seeking to minimise the modular description length for random walks on networks:
Without communities, that is, when all nodes are assigned to the same module, the codelength is simply the entropy over the nodes' visit rates as formalised in \Cref{eq:entropy}.
For a two-level partition, the map equation calculates the random walk's per-step description length $L$ -- also called \emph{codelength} -- as a weighted average of the modules' entropies and the entropy of a so-called index-level module for switching between modules,
\begin{equation}
    L\left(\mathsf{M}\right) = q H\left(Q\right) + \sum_{\mathsf{m} \in \mathsf{M}} p_\mathsf{m} H\left(P_\mathsf{m}\right).
\end{equation}
Here $\mathsf{M}$ is a partition of the nodes into modules, $q = \sum_{\mathsf{m} \in \mathsf{M}} q_\mathsf{m}$ is the index-level codebook usage rate, $q_\mathsf{m}$ is the entry rate for module $\mathsf{m}$, $Q = \left\{ q_\mathsf{m} \,|\, \mathsf{m} \in \mathsf{M} \right\}$ is the set of module entry rates, $p_\mathsf{m} = \mathsf{m}_\text{exit} + \sum_{u \in \mathsf{m}} p_u$ is module $\mathsf{m}$'s codebook usage rate, $\mathsf{m}_\text{exit}$ is module $\mathsf{m}$'s exit rate, and $P_\mathsf{m} = \left\{ \mathsf{m}_\text{exit} \right\} \cup \left\{ p_u \,|\, u \in \mathsf{m} \right\}$ is the set of node visit rates in module $\mathsf{m}$, including its exit rate.
\Cref{fig:map-equation-principle}b shows an example for a two-level coding scheme where codewords are reused across modules for a shorter overall codelength.
Through recursion, the map equation generalises to hierarchical partitions \cite{rosvall2011multilevelmapeq,map-equation-review}.

Minimising the map equation creates a \emph{map} of the network's organisational structure where nodes are grouped into modules such that a random walker tends to stay within modules and switching between modules occurs rarely.
We can draw such maps as a tree as shown in \Cref{fig:map-equation-principle}c.
Each random-walker step along a link in the network corresponds to traversing the map along the shortest path between these two nodes; to describe the step, we use the codewords along the path in the map.

To apply the ideas behind the KL divergence, we rewrite the map equation and express it as a random walk, matching the form of \Cref{eq:as-modular-random-walk},
\begin{align}
    L\left(\mathsf{M}\right)
    & = q H\left(Q\right) + \sum_{\mathsf{m} \in \mathsf{M}} p_\mathsf{m} H\left(P_\mathsf{m}\right) \nonumber \\
    & = - \sum_{u \in V} p_u \sum_{v \in V} t_{uv} \log_2 \operatorname{mapsim}\left(\mathsf{M}, u, v\right),
    \label{eq:map-equation-rewrite}
\end{align}
where $\log_2 \operatorname{mapsim}\left(\mathsf{M}, u, v\right)$ is the number of bits required to describe a random-walker step from node $u$ to $v$, given partition $\mathsf{M}$~\cite{pmlr-v198-blocker22a}.
For two-level partitions, $\operatorname{mapsim}$, which is short for map equation similarity, is defined as
\begin{align}
    \operatorname{mapsim}\left(\mathsf{M}, u, v\right) \! = \!
    \begin{dcases*}
        \frac{p_v}{p_{\mathsf{m}_v}}
           & if $\mathsf{m}_u \! = \! \mathsf{m}_v$, \\
        \frac{\mathsf{m}_{u,\text{exit}}}{p_{\mathsf{m}_u}} \!\cdot\! \frac{q_{\mathsf{m}_v}}{q} \!\cdot\! \frac{p_v}{p_{\mathsf{m}_v}} \!\!\!
           & if $\mathsf{m}_u \! \neq \! \mathsf{m}_v$,
    \end{dcases*}
    \label{eq:mapsim}
\end{align}
where, $\mathsf{m}_u$ and $\mathsf{m}_v$ are the modules to which nodes $u$ and $v$ belong, respectively.
Based on a network's map, $\operatorname{mapsim}$ quantifies the rate at which a random walker transitions between pairs of nodes.
Importantly, $\operatorname{mapsim}$ depends on the source node's module, but not on the source node itself.
We detail the derivation of \Cref{eq:map-equation-rewrite} in \Cref{appx:derivation}.

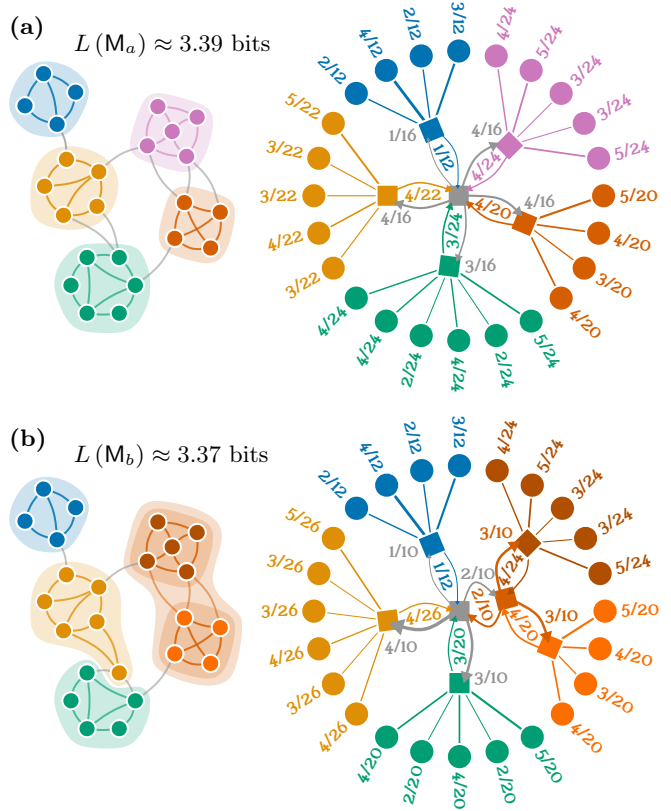
\begin{figure}[ht!]
    \centering
    \begin{overpic}[width=1\linewidth]{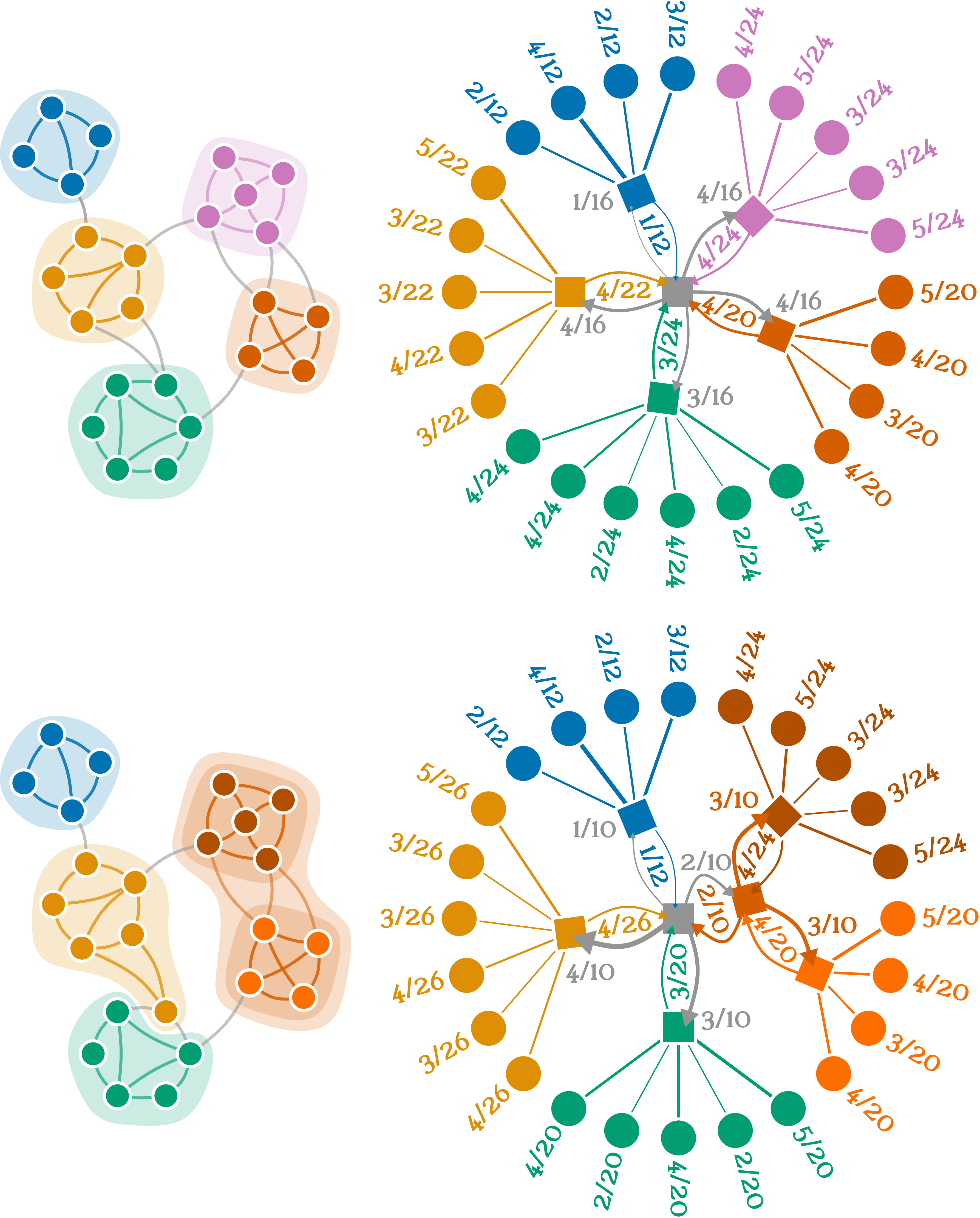}
        \put(0,96){\textbf{(a)}}
        \put(8,94){$L\left(\mathsf{M}_a\right) \approx 3.39$ bits}
        \put(0,45){\textbf{(b)}}
        \put(8,43){$L\left(\mathsf{M}_b\right) \approx 3.37$ bits}
    \end{overpic}
    \caption{Different partitions for the same network, drawn on the network in the left column and as a tree in the right column.
    \textbf{(a)} A two-level partition of the network into five modules.
    \textbf{(b)} A three-level partition of the network into four modules, one of which has two submodules.
    Labels in the trees show the rate at which a random walker visits nodes and enters or exits modules.
    For example, a random walker who is in the blue module exits at rate $\frac{1}{12}$.
    A random walker who is at the tree's root level enters the green module at rate $\frac{3}{24}$ in partition \textbf{(a)} and $\frac{3}{20}$ in partition \textbf{(b)}, respectively.
    }
    \label{fig:comparing-maps}
\end{figure}

\subsection{Relative Entropy Between Two Maps}
Essentially, a map summarises the random walker's movement pattern on a network and represents that pattern as an ensemble of possibly nested random processes.
Different maps of the same network summarise the patterns differently, resulting in different codelengths.

Taking inspiration from the KL divergence, we make a first attempt to define our dissimilarity measure for comparing two maps $\mathsf{M}_a$ and $\mathsf{M}_b$,
\begin{align}
    D \left( \mathsf{M}_a \,||\, \mathsf{M}_b \right) \! = \! \sum_{u \in V} \! p_u \! \sum_{v \in V} \! t_{uv} \log_2 \frac{\operatorname{mapsim}\left(\mathsf{M}_a, u, v\right)}{\operatorname{mapsim}\left(\mathsf{M}_b, u, v\right)}.
    \label{eqn:flow-divergence-naive}
\end{align}
However, using the transition rates $t_{uv}$ only computes the codelength difference between two partitions:
By simplifying \Cref{eqn:flow-divergence-naive} and substituting with \Cref{eq:map-equation-rewrite}, we get $D\left(\mathsf{M}_a \,||\, \mathsf{M}_b\right) = L\left(\mathsf{M}_a\right) - L\left(\mathsf{M}_b\right)$.
For the partitions shown in \Cref{fig:comparing-maps}, we get $D\left(\mathsf{M}_a \,||\, \mathsf{M}_b\right) = 0.02~\text{bits}$ and $D\left(\mathsf{M}_b \,||\, \mathsf{M}_a\right) = -0.02~\text{bits}$.
That is, we would simply learn by how much the codelength would increase or decrease, were we to use the other partition for describing the random walk.

But this is not what we intend to measure.
Following the idea of the KL divergence, we assume that $\mathsf{M}_a$ captures the random walker's true movement patterns.
And we ask: what is the expected additional number of bits required to describe a random walk using an estimate $\mathsf{M}_b$ of its true patterns $\mathsf{M}_a$?
Therefore, we use the partition-dependent transition rates $t_{uv}^{\mathsf{M}_a}$ and define our partition dissimilarity measure \emph{flow divergence},
\begin{align}
    \!\!\! D_F \! \left( \mathsf{M}_a \,||\, \mathsf{M}_b \right) \! = \!\! \sum_{u \in V} \! p_u \!\! \sum_{v \in V} \! t_{uv}^{\mathsf{M}_a} \log_2 \! \frac{\operatorname{mapsim}\left(\mathsf{M}_a, u, v\right)}{\operatorname{mapsim}\left(\mathsf{M}_b, u, v\right)}.\!
    \label{eqn:flow-divergence}
\end{align}

\subsection{Walking on Maps and Deriving Transition Rates}
When we have obtained a map that describes a network's organisational structure, for example by minimising the map equation or using any other community-detection method, we use \Cref{eq:mapsim} to derive $\operatorname{mapsim}$ values between all pairs of nodes.

\begin{figure}[t!]
    \centering
    \begin{overpic}[width=1\linewidth]{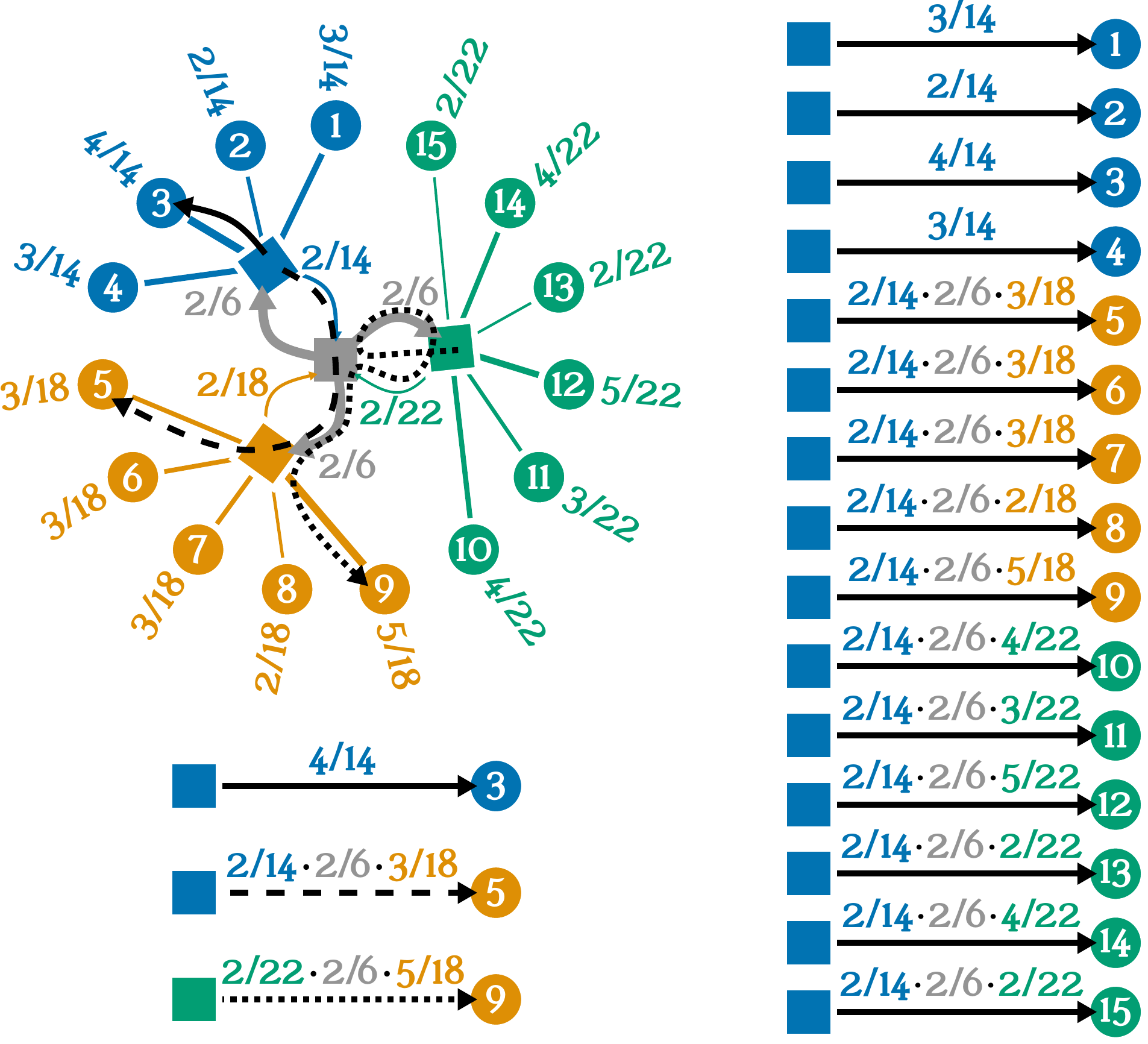}
        \put(0,90){\textbf{(a)}}
        \put(5,25){\textbf{(b)}}
        \put(60,90){\textbf{(c)}}
    \end{overpic}
    \caption{Walking on maps.
    \textbf{(a)} The same map as in \Cref{fig:map-equation-principle}c, but now annotated with module-normalised node visit rates instead of codewords. The solid, dashed, and dotted arrows show examples of three random walker paths on the map.
    \textbf{(b)} We derive transition rates between pairs of nodes according to $\operatorname{mapsim}$: Transition rates depend on the source node's module, not on the source node itself (\cref{eq:mapsim}). Therefore, the shortest paths on the map start at module nodes and we obtain the rate at which each shortest path is used by multiplying the transition rates along that path. The dotted arrow does not represent a shortest path because it contains a loop which we need to remove.
    \textbf{(c)} All shortest paths that start in the blue module and their usage rates.
    }
    \label{fig:walking-on-maps}
\end{figure}

Consider the map shown in \Cref{fig:walking-on-maps}a where nodes are annotated with their module-normalised visit rates and arrows between modules show the modules' entry and exit rates.
There are also three black arrows drawn on the map: a solid arrow for a random walker who is in the blue module and steps to node $3$, a dashed arrow for a random walker who is in the blue module and steps to node $5$ in the orange module, and a dotted arrow for a random walker who is in the green module and visits node $9$ in the orange module.
Because codewords depend on the random walker's current module, but not on the current node, the shortest paths in the map begin at the module nodes.

\Cref{fig:walking-on-maps}b details at what rates a random walker uses the solid, dashed, and dotted paths in the map.
A random walker who is in the blue module visits node $3$ at rate $\frac{4}{14}$.
A random walker who is in the blue module exits at rate $\frac{2}{14}$, then enters the orange module at rate $\frac{2}{6}$, and then visits node $5$ at rate $\frac{3}{18}$, resulting in a rate of $\frac{2}{14} \cdot \frac{2}{6} \cdot \frac{3}{18}$ for the dashed arrow.
The dotted arrow contains a loop, which we discard because we describe transitions along the shortest paths in the map.
Therefore, a random walker who is in the green module exits at rate $\frac{2}{22}$, then enters the orange module at rate $\frac{2}{6}$, and visits node $9$ at rate $\frac{5}{18}$, resulting in a rate of $\frac{2}{22} \cdot \frac{2}{6} \cdot \frac{5}{18}$ for the dotted arrow.
Taking the $\log_2$ of the arrows' usage rates returns the required number of bits for describing the corresponding random-walker step.

\Cref{fig:walking-on-maps}c shows all shortest paths that start in the blue module together with their usage rates.
However, since we only consider shortest paths in the map and not paths that contain loops, their usage rates do not sum to 1; here the rates for shortest paths starting in the blue module sum to approximately $0.94$.
Because paths with loops are ways to make detours from shortest paths and returning to them, their usage rate is proportional to that of their contained shortest path.
To obtain the transition probability from $u$ to $v$ according to partition $\mathsf{M}$, we normalise with the sum of transition rates for shortest paths from $u$ to all nodes $v$, except for $u$ itself because we assume no self-links,
\begin{equation}
    t_{uv}^\mathsf{M} = \frac{\operatorname{mapsim}\left(\mathsf{M}, u, v\right)}{\sum_{v \left(\neq u\right)} \operatorname{mapsim}\left(\mathsf{M}, u, v\right)}.
    \label{eq:module-dependent-transition-rates}
\end{equation}
If we want to consider self-links, we also include $u$ itself.

Comparing the maps shown in \Cref{fig:comparing-maps},
we obtain $D_F\left(\mathsf{M}_a \,||\, \mathsf{M}_b\right) \approx 0.11$~\text{bits} and $D_F\left(\mathsf{M}_b \,||\, \mathsf{M}_a\right) \approx 0.01$~\text{bits}.
This means that, assuming that $\mathsf{M}_a$ captures the network's true community structure, the expected additional cost per random-walker step for using $\mathsf{M}_b$ is $0.11$~bits.
Conversely, assuming that $\mathsf{M}_b$ describes the network's true structure, the expected additional cost per step for using $\mathsf{M}_a$ is $0.01$~bits.

\section{Results}
We apply flow divergence to compute partition dissimilarity scores for partitions in synthetic and real networks.
In the synthetic network, we show that flow divergence can distinguish between partitions that popular measures consider equally good compared to a reference partition.
In real networks, we use flow divergence to quantify the cost of overfitting in incomplete data and to compute dissimilarity matrices for embedding partitions.

\subsection{Synthetic Example}\label{sec:synthetic}
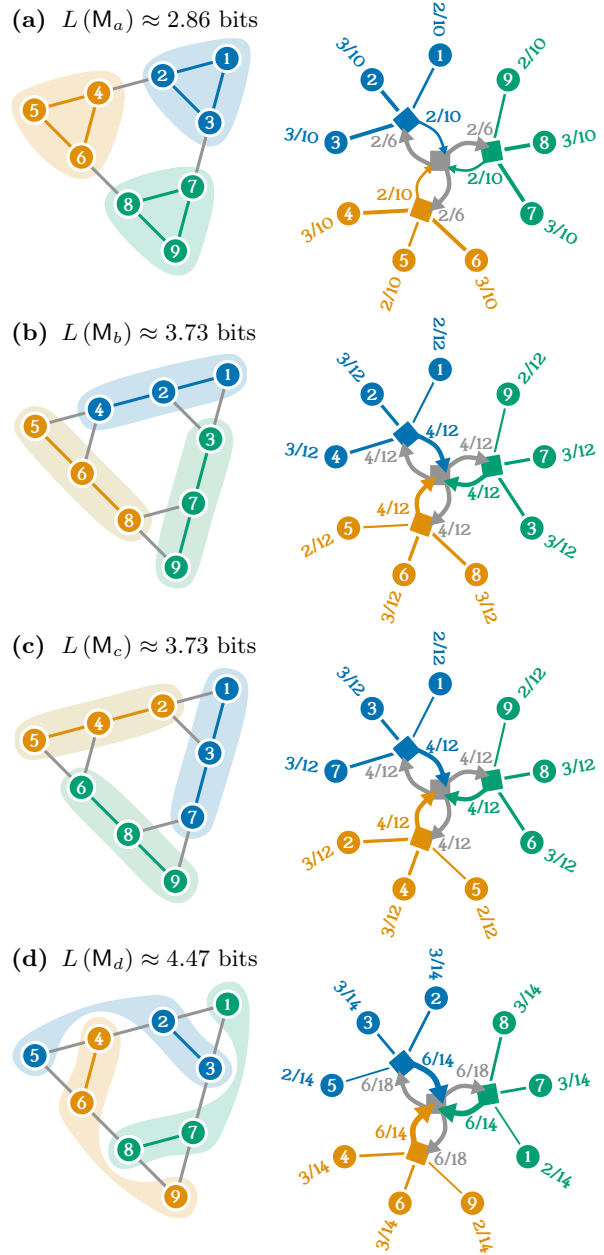
\begin{figure}[ht!]
    \centering
    \begin{overpic}[width=.9\linewidth]{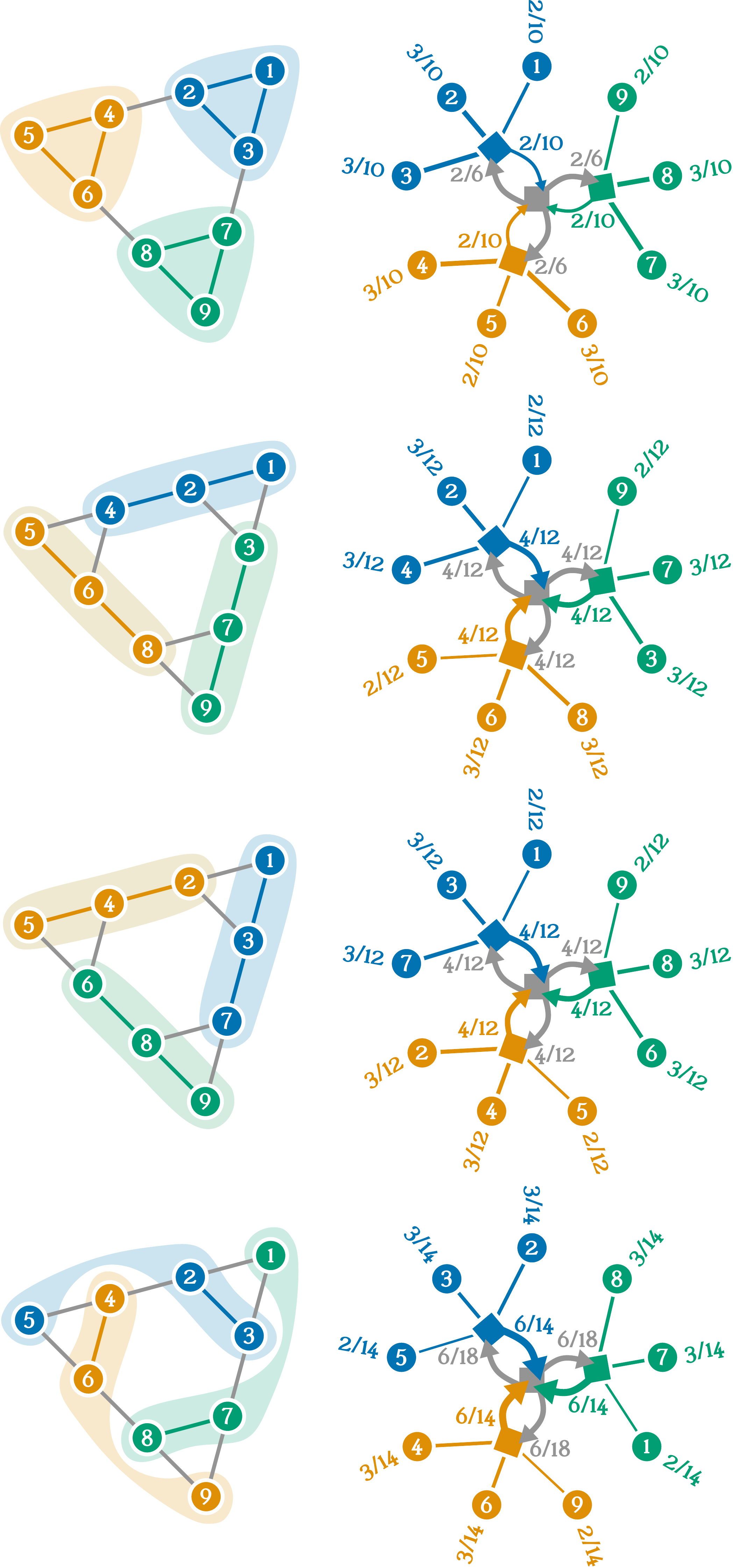}
        \put(0,98){\textbf{(a)}}
        \put(4,98){$L\left(\mathsf{M}_a\right) \approx 2.86$ bits}
        \put(0,73){\textbf{(b)}}
        \put(4,73){$L\left(\mathsf{M}_b\right) \approx 3.73$ bits}
        \put(0,48){\textbf{(c)}}
        \put(4,48){$L\left(\mathsf{M}_c\right) \approx 3.73$ bits}
        \put(0,23){\textbf{(d)}}
        \put(4,23){$L\left(\mathsf{M}_d\right) \approx 4.47$ bits}
    \end{overpic}
    \caption{Comparing maps. The same partitions as shown in \Cref{fig:shortcomings}, together with their maps.
    Flow divergence can distinguish between partitions that popular measures, such as the Jaccard index and mutual information, consider equally good with respect to a reference partition.
    \textbf{(a)} The reference partition $\mathsf{M}_a$.
    The partitions $\mathsf{M}_b$ in \textbf{(b)} and $\mathsf{M}_c$ in \textbf{(c)} have the same codelength and are symmetric: each module overlaps in two out of three nodes with the modules in the reference partition.
    \textbf{(d)} Partition $\mathsf{M}_d$ with disconnected communities but still a two-out-of-three overlap per module with the reference partition.
    }
    \label{fig:motivation-with-maps}
\end{figure}

We return to our initial example and confirm that flow divergence can distinguish between partitions where the Jaccard index and mutual information cannot.
\Cref{fig:motivation-with-maps} shows the networks and their maps, annotated with transition rates.
We use \Cref{eqn:flow-divergence} to compute dissimilarities and list them in \Cref{tab:motivation-divergences}.
Despite its higher codelength, $\mathsf{M}_d$ is more similar to $\mathsf{M}_a$ than $\mathsf{M}_b$ and $\mathsf{M}_c$ are.
This highlights an important insight: better maps do not necessarily have a lower flow divergence.

\begin{table}[h!]
 \caption{Flow divergence in bits, rounded to two decimal places, between the partitions shown in \Cref{fig:motivation-with-maps}.}
  \centering
  \begin{tabular}{ccccc}
    \toprule
    & \multicolumn{4}{c}{Other} \\
    \cmidrule(r){2-5}
    Reference & $\mathsf{M}_a$ & $\mathsf{M}_b$ & $\mathsf{M}_c$ & $\mathsf{M}_d$ \\
    \midrule
    $\mathsf{M}_a$ & 0    & 1.92 & 1.92 & 1.5  \\
    $\mathsf{M}_b$ & 1.8  & 0    & 1.17 & 1.99 \\
    $\mathsf{M}_c$ & 1.8  & 1.17 & 0    & 1.48 \\
    $\mathsf{M}_d$ & 1.14 & 1.78 & 1.25 & 0    \\
    \bottomrule
  \end{tabular}
  \label{tab:motivation-divergences}
\end{table}

For a more complete picture and to explain why $\mathsf{M}_d$ is closer to $\mathsf{M}_a$, we consider the individual nodes' contribution to the divergences.
Instead of computing the outer sum in \Cref{eqn:flow-divergence}, we report flow divergence values per node in \Cref{tab:motivation-divergences-individual}.
Interpreting these results, we can explain that $\mathsf{M}_d$ is more similar to $\mathsf{M}_a$ because its modules overlap in the higher-degree nodes.
In contrast, the modules in $\mathsf{M}_b$ and $\mathsf{M}_c$ overlap in one higher and one lower-degree node with the modules in $\mathsf{M}_a$.
Therefore, with respect to flow, $\mathsf{M}_d$ is more similar to $\mathsf{M}_a$ than $\mathsf{M}_b$ and $\mathsf{M}_c$ are.

\begin{table}[h!]
 \caption{Flow divergence on a per-node basis in bits, rounded to two decimal places, between the partitions shown in \Cref{fig:motivation-with-maps} where we fix $\mathsf{M}_a$ as the reference partition.}
  \centering
  \begin{tabular}{cccccccccc}
    \toprule
    & \multicolumn{9}{c}{Node} \\
    \cmidrule(r){2-10}
    Other & 1 & 2 & 3 & 4 & 5 & 6 & 7 & 8 & 9 \\
    \midrule
    $\mathsf{M}_b$ & 0.12 & 0.2  & 0.32 & 0.32 & 0.12 & 0.2  & 0.2  & 0.32 & 0.12 \\
    $\mathsf{M}_c$ & 0.12 & 0.32 & 0.2  & 0.2  & 0.12 & 0.32 & 0.32 & 0.2  & 0.12 \\
    $\mathsf{M}_d$ & 0.21 & 0.14 & 0.14 & 0.14 & 0.21 & 0.14 & 0.14 & 0.14 & 0.21 \\
    \bottomrule
  \end{tabular}
  \label{tab:motivation-divergences-individual}
\end{table}

\subsection{The Cost of Overfitting}
Real-world data is often incomplete, resulting in spurious patterns in observed networks.
Community-detection methods may pick up communities that exist purely due to chance because the data is incomplete~\cite{smiljanic2021comnet}.
In the case of the map equation, less data generally reduces the codelength: fewer links make networks sparser, leading to smaller modules with lower entropy.
However, the process that guides how links form does not change and we have already seen in \Cref{sec:synthetic} that lower-codelength partitions do not necessarily capture the ``true'' dynamics best.

Here, we consider the cost of overfitting due to incomplete data.
We choose four real-world social networks (\Cref{tab:networks}) and use Infomap~\cite{edler2017mapequation,mapequation2022software}, the map equation's optimisation algorithm, to detect communities.
We use those communities as the reference partition.
Then, we remove an $r$-fraction of the links and use Infomap to detect communities in the reduced network.
For removing links, we consider them in random order, removing one at a time until we have removed an $r$-fraction.
However, if removing a link would split the network into disconnected components, we keep the link and continue with the next one.
The reduced network contains at least $\left|V\right|-1$ many links, placing an upper bound on the $r$-fraction we can remove.
Finally, we use flow divergence to compute the expected additional cost in bits for overfitting, that is, relative to the reference partition based on ``complete'' data.
For each $r$, we repeat this $100$ times and report averages.

\begin{figure}[t!]
    \centering
    \vspace{\baselineskip}
    \begin{overpic}[width=.49\linewidth]{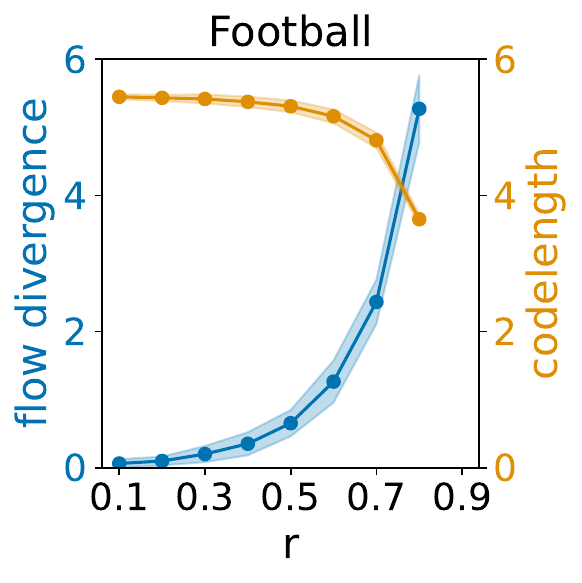}
        \put(0,100){\textbf{(a)}}
    \end{overpic}
    \begin{overpic}[width=.49\linewidth]{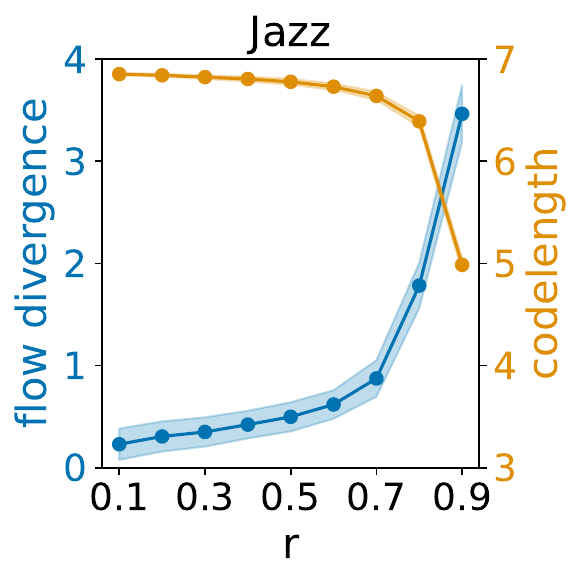}
        \put(0,100){\textbf{(b)}}
    \end{overpic}

    \vspace{\baselineskip}
    \begin{overpic}[width=.49\linewidth]{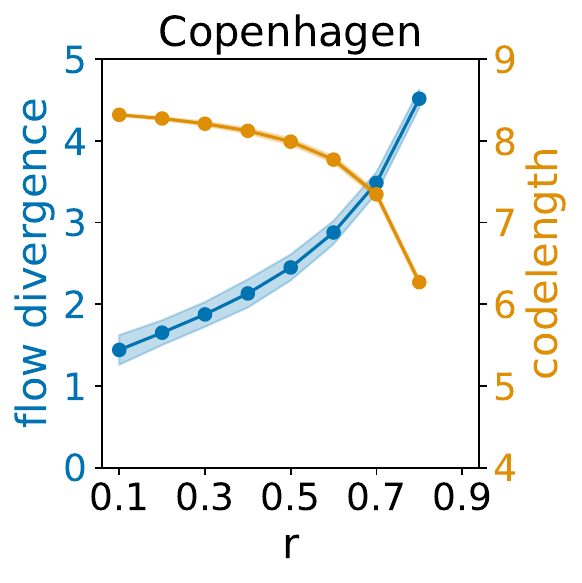}
        \put(0,100){\textbf{(c)}}
    \end{overpic}
    \begin{overpic}[width=.49\linewidth]{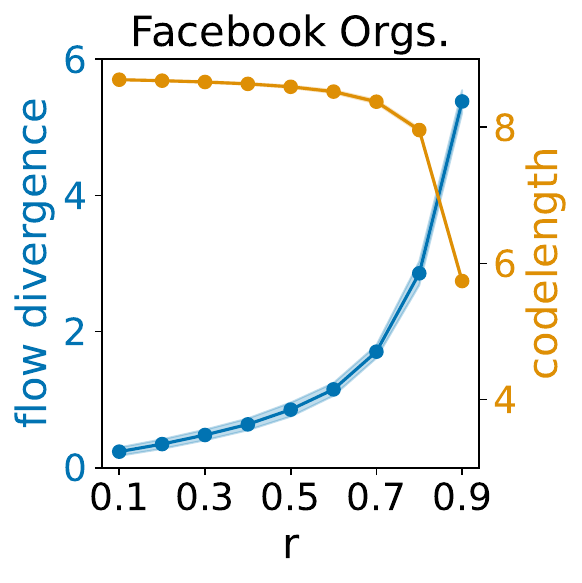}
        \put(0,100){\textbf{(d)}}
    \end{overpic}
    \caption{The cost of overfitting.
    We use real-world networks and remove different $r$-fractions of their links and partition the reduced network with Infomap.
    We measure the detected partitions' divergence against the reference partition obtained from the complete network.
    For each value of $r$, we repeat the sampling 100 times and report averages; the error band show one standard deviation.
    We find that, as $r$ increases, flow divergence increases faster than the codelength decreases.
    \textbf{(a)} A network of college football clubs that have played against each other.
    \textbf{(b)} A collaboration network between jazz musicians.
    \textbf{(c)} A social network between students.
    \textbf{(d)} A social network between the employees of a company.
    }
    \label{fig:cost-of-overfitting}
\end{figure}

We find that, as $r$ increases, flow divergence increases faster than the codelength decreases (\Cref{fig:cost-of-overfitting}).
While the description of random walks on the network can be compressed more when less data is available, flow divergence tells us about the actual cost of using what seems like a more efficient encoding.
In other words: with incomplete data, Infomap finds modules that enable encoding random walks more efficiently.
However, as flow divergence shows, these modules diverge more from what we assume to be the true dynamics on the network.

\begin{table}[hb!]
 \caption{Properties of four real networks that we use for measuring the cost of overfitting. We list the number of nodes, $\left|V\right|$, the number of links $\left|E\right|$, average degree $\left\langle k \right\rangle$, and the codelength $L$ for the reference partition detected by Infomap.}
  \centering
  \begin{tabular}{lccccc}
    \toprule
    Network & Ref. & $\left|V\right|$ & $\left|E\right|$ & $\left\langle k \right\rangle$ & $L$ \\
    \midrule
    Football            & \cite{doi:10.1073/pnas.122653799}    &   115 &    613 & 10.7 & 5.45 \\
    Jazz                & \cite{doi:10.1142/S0219525903001067} &   198 &  2,742 & 27.7 & 6.86 \\
    Copenhagen          & \cite{Sapiezynski2019}               &   800 &  6,429 & 16.1 & 8.34 \\
    Facebook Orgs.      & \cite{Fire2016}                      & 1,429 & 19,357 & 27.1 & 8.71 \\
    \bottomrule
  \end{tabular}
  \label{tab:networks}
\end{table}

\subsection{Embedding Partitions}
Flow divergence can be used to compute dissimilarity matrices for visualising a network's solution landscape.
We consider the Jazz network which is known to have a diverse solution landscape~\cite{map-equation-review}.
We run Infomap 200 times with different seeds to obtain 200 partitions and use flow divergence to compute all pairwise distances between those 200 partitions.
For comparison, we also compute dissimilarity matrices using the Jaccard index and AMI.
Then, we use UMAP~\cite{mcinnes2020umap} to embed the partitions in 2D, setting the number of neighbours to 100 and the minimum distance to 1.
We use HDBSCAN~\cite{10.1007/978-3-642-37456-2_14} to identify clusters in flow divergence embedding space and mark those clusters in all three embeddings with colours.

\begin{figure*}[ht!]
    \centering
    \vspace{1.5\baselineskip}
    \begin{overpic}[width=\linewidth]{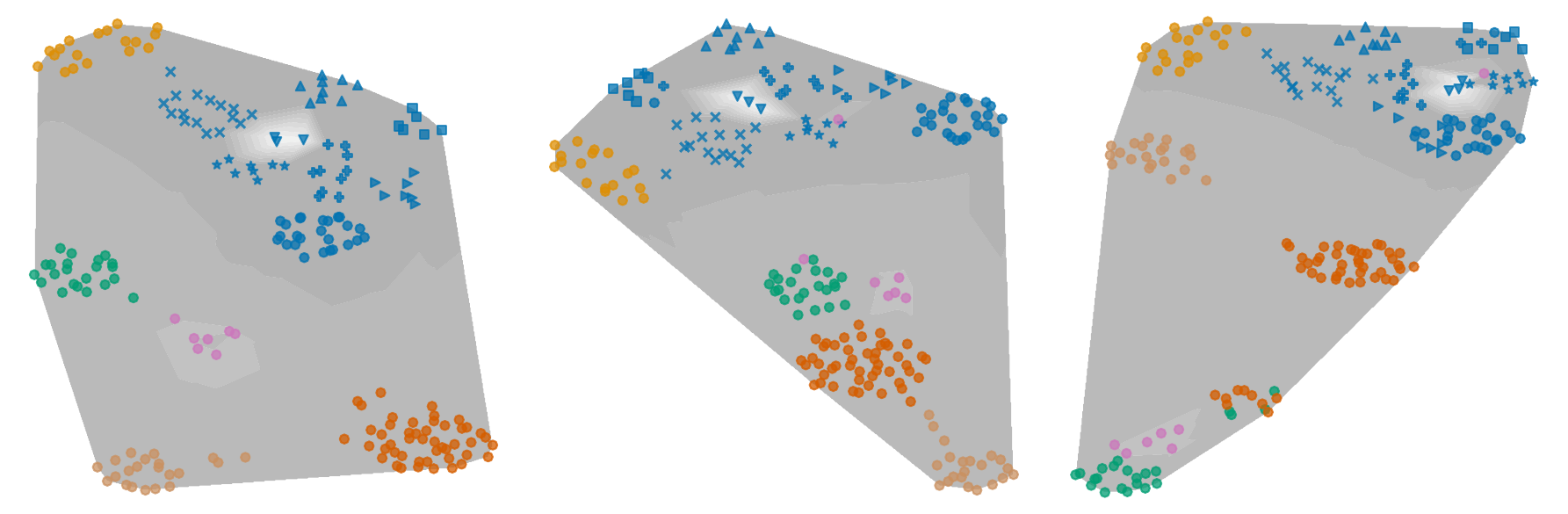}
        \put(0,33){\textbf{(a)}}
        \put(34,33){\textbf{(b)}}
        \put(68,33){\textbf{(c)}}
    \end{overpic}
    \caption{Embeddings of 200 Infomap partitions for the Jazz network where lighter regions correspond to a higher codelength.
    We use UMAP to embed the partitions based on their dissimilarities according to \textbf{(a)} flow divergence, \textbf{(b)} the Jaccard index, and \textbf{(c)} adjusted mutual information, and set the number of neighbours to 100 and the minimum distance to 1.
    We detect clusters of partitions in flow divergence embedding space with HDBSCAN and colour the resulting groups.
    The blue partitions are somewhat less separated from each other and we use shapes to highlight the visually identified sub-groups.
    Overall, the three measures produce similar embeddings, however, flow divergence separates the red and green groups more clearly; the Jaccard index and AMI consider the pink group more similar to the green group and mix one pink partition into the blue group; the Jaccard index and AMI mix some of the blue sub-groups that flow divergence separates more distinctly.
    }
    \label{fig:embeddings}
\end{figure*}

\Cref{fig:embeddings} shows the Jazz network's solution landscapes.
We find that flow divergence distinguishes clearly between partitions that the Jaccard index and adjusted mutual information consider more similar.
While flow divergence separates the red and green groups, the Jaccard index places them relatively close together.
AMI separates the red and green groups but also identifies an additional group that is a mix of red and green partitions.
All three measures place the pink group near the green group, however, the Jaccard index also considers it more similar to the red group than flow divergence and AMI.
Interestingly, both the Jaccard index and AMI place one of the pink partitions relatively centrally in the blue group.
The blue group contains several sub-groups which we identify visually in flow divergence embedding space and plot with different marker shapes.
The Jaccard index and AMI agree with flow divergence regarding some of the blue sub-groups but mix squares, circles, and crosses to some extent.

However, if we were to determine groups based on Jaccard or AMI embedding space, we would see a similar picture where the other measures mix their groups.
Similar to how no community-detection method is fundamentally more correct than any other, we cannot say that a particular partition similarity score is always the best.
Which measure is the most appropriate depends on the context and should be selected based on the research question.

\subsection{Computational Complexity and Limitations}
Computing flow divergence between two partitions requires considering $n^2$ $\operatorname{mapsim}$ scores per partition, where $n = \left|V\right|$.
Because $\operatorname{mapsim}\left(\mathsf{M}, u, v\right)$ depends on the source node's module, $\mathsf{m}_u$, but not the source node $u$ itself, we obtain the same values for different source nodes in the same module when the target node $v$ is the same.
Therefore, the number of $\operatorname{mapsim}$ values we need to compute reduces to $m \cdot n$, where $m \ll n$ is the number of modules, typically scaling as $\sqrt{n}$~\cite{8692626}.
For each $\operatorname{mapsim}$ value, we need to find the source module $\mathsf{m}_u$ and target node $v$ in the partition tree and multiply the random walker's transition rates along the shortest path from $\mathsf{m}_u$ to $v$, which can be done in time $\mathcal{O}\left(\log n\right)$.
For all $m \cdot n$ $\operatorname{mapsim}$ values, this requires time $\mathcal{O}\left(m \cdot n \cdot \log n\right) = \mathcal{O}\left(\sqrt{n} \cdot n \cdot \log n\right) \in \mathcal{O}\left(n^2\right)$.
Since we compute partition-dependent transition rates between all pairs of nodes (\Cref{eq:module-dependent-transition-rates}), the overall complexity is $\mathcal{O}\left(n^2\right)$.

Because flow divergence is based on the KL divergence, similar limitations apply:
To compare two probability distributions, $P$ and $Q$ that are defined on the same sample space $X$, the KL divergence (\Cref{eq:kl-divergence}) is only defined if, for all $x \in X$, $q_x = 0$ implies $p_x$ = 0.
Otherwise, $D\left(P \,||\, Q\right) = \infty$.
For flow divergence, this means that we require networks to be connected:
$\operatorname{mapsim}$ values between nodes in disconnected parts of the network become $0$, resulting in transitions with probability $0$, which in turn would yield an infinitely high flow divergence.

To handle disconnected networks, we could add a small constant to each $\operatorname{mapsim}$ value, thus ensuring that all transition probabilities are larger than 0.
Alternatively, we could regularise the random walker's transition rates with a Bayesian prior that was designed to avoid overfitting~\cite{smiljanic2021comnet}.
However, both approaches would increase the codelength because they add further links to the network.

\section{Conclusion}
We have studied the problem of comparing network partitions and motivated the need for approaches that take link patterns into account: while community detection focuses on grouping nodes together that share similar link patterns, measures for comparing partitions typically ignore links altogether.

Inspired by the Kullback-Leibler divergence for measuring ``distances'' between probability distributions, we developed a partition dissimilarity score based on random walks: \emph{flow divergence}.
Flow divergence is related to the map equation and quantifies the expected additional number of bits per step for describing a random walk on a network when using an estimate $\mathsf{M}_b$ of the network's actual community structure $\mathsf{M}_a$.

Applied to synthetic networks, we have shown that, by incorporating link patterns, flow divergence can distinguish between partitions that popular partition similarity measures consider to be equally good.
In real networks, we have highlighted how flow divergence gives insights into the cost of overfitting when detecting communities in a network whose data is incomplete.
Moreover, we have used flow divergence to compute distance matrices between partitions for creating embedding with UMAP to visualise a network's solution landscape.

\section*{Acknowledgments}
We thank Martin Rosvall and Daniel Edler for helpful discussions.
Christopher Bl{\"o}cker and Ingo Scholtes acknowledge funding from the Swiss National Science Foundation, grant 176938, and the German Federal Ministry of Education and Research, grant 100582863 (TissueNet).

\bibliographystyle{unsrtnat}

\onecolumngrid
\appendix
\section{Rewriting the Map Equation}\label{appx:derivation}
Let $G = \left(V, E, \delta\right)$ be a network with nodes $V$, links $E$, and links weights $\delta\colon E \to \mathbb{R}^+$.
If $G$ is undirected, we can calculate the stationary visit rate for node $u$ as $p_u = \frac{\sum_{v \in V} \delta\left(u, v\right)}{\sum_{u \in V} \sum_{v \in V} \delta\left(u,v\right)}$.
If $G$ is directed, we can use a power iteration to solve the recursive set of equations $p_v = \sum_{u \in V} p_u t_{uv}$, where $t_{uv}$ is the probability that a random walker at $u$ steps to $v$.

We begin with the two-level map equation \cite{doi:10.1073/pnas.0706851105},
\begin{align*}
    L\left(\mathsf{M}\right) & = q H\left(Q\right) + \sum_{\mathsf{m} \in \mathsf{M}} p_\mathsf{m} H\left(P_\mathsf{m}\right),
\end{align*}
where $\mathsf{M}$ is a partition of the nodes into modules, $q = \sum_{\mathsf{m} \in \mathsf{M}} q_\mathsf{m}$ is the index-level codebook usage rate, $q_\mathsf{m}$ is the entry rate for module $\mathsf{m}$, $Q = \left\{ q_\mathsf{m} \,|\, \mathsf{m} \in \mathsf{M} \right\}$ is the set of module entry rates, $p_\mathsf{m} = \mathsf{m}_\text{exit} + \sum_{u \in \mathsf{m}} p_u$ is module $\mathsf{m}$'s codebook usage rate, $\mathsf{m}_\text{exit}$ is module $\mathsf{m}$'s exit rate, and $P_\mathsf{m} = \left\{ \mathsf{m}_\text{exit} \right\} \cup \left\{ p_u \,|\, u \in \mathsf{m} \right\}$ is the set of node visit rates in module $\mathsf{m}$, including its exit rate.

Expanding the map equation, we obtain
\begin{align*}
    - q \sum_{\mathsf{m} \in \mathsf{M}} \frac{q_\mathsf{m}}{q} \log_2 \frac{q_\mathsf{m}}{q}
    - \sum_{\mathsf{m} \in \mathsf{M}} p_\mathsf{m} \left( \frac{\mathsf{m}_\text{exit}}{p_\mathsf{m}} \log_2 \frac{\mathsf{m}_\text{exit}}{p_\mathsf{m}}
    + \sum_{u \in \mathsf{m}} \frac{p_u}{p_\mathsf{m}} \log_2 \frac{p_u}{p_\mathsf{m}} \right),
\end{align*}
and after cancelling common factors
\begin{align*}
    - \sum_{\mathsf{m} \in \mathsf{M}} q_\mathsf{m} \log_2 \frac{q_\mathsf{m}}{q}
    - \sum_{\mathsf{m} \in \mathsf{M}} \left( \mathsf{m}_\text{exit} \log_2 \frac{\mathsf{m}_\text{exit}}{p_\mathsf{m}} + \sum_{u \in \mathsf{m}} p_u \log_2 \frac{p_u}{p_\mathsf{m}} \right).
\end{align*}
Pulling out the summation and annotating the parts, we have
\begin{align*}
    - \sum_{\mathsf{m} \in \mathsf{M}}
    \overbracket{q_\mathsf{m} \log_2 \frac{q_\mathsf{m}}{q}}^{\text{entering module } \mathsf{m}}
    + \overbracket{\mathsf{m}_\text{exit} \log_2 \frac{\mathsf{m}_\text{exit}}{p_\mathsf{m}}}^{\text{exiting module } \mathsf{m}}
    + \overbracket{\sum_{u \in \mathsf{m}} p_u \log_2 \frac{p_u}{p_\mathsf{m}}.}^{\text{visiting nodes in module } \mathsf{m}}
\end{align*}
Next, we use $q_\mathsf{m} = \sum_{u \not\in \mathsf{m}} p_u \sum_{v \in \mathsf{m}} t_{uv}$, $\mathsf{m}_\text{exit} = \sum_{u \in \mathsf{m}} p_u \sum_{v \not\in \mathsf{m}} t_{uv}$, and $p_v = \sum_u p_u t_{uv}$, and split up the last part,
\begin{align*}
    - \sum_{\mathsf{m} \in \mathsf{M}} & \left( \sum_{u \not\in \mathsf{m}} p_u \sum_{v \in \mathsf{m}} t_{uv} \log_2 \frac{q_\mathsf{m}}{q} \right)
    + \left( \sum_{u \in \mathsf{m}} p_u \sum_{v \not\in \mathsf{m}} t_{uv} \log_2 \frac{\mathsf{m}_\text{exit}}{p_\mathsf{m}} \right) \\
    & + \left( \sum_{u \in \mathsf{m}} p_u \sum_{v \in \mathsf{m}} t_{uv} \log_2 \frac{p_v}{p_\mathsf{m}} \right)
    + \left( \sum_{u \not\in \mathsf{m}} p_u \sum_{v \in \mathsf{m}} t_{uv} \log_2 \frac{p_v}{p_\mathsf{m}} \right).
\end{align*}
Then, we merge the second and fourth term into the first term.
To merge the second term, we turn the module exits around, considering those steps that leave other modules to enter module $\mathsf{m}$ instead of steps that leave module $\mathsf{m}$.
We denote node $u$'s and $v$'s module by $\mathsf{m}_u$ and $\mathsf{m}_v$, respectively,
\begin{align*}
    - \sum_{\mathsf{m} \in \mathsf{M}} \Biggl( \sum_{u \not\in \mathsf{m}} p_u \sum_{v \in \mathsf{m}} t_{uv} \log_2 \Bigl( \overbracket{\frac{q_{\mathsf{m}_u}}{p_{\mathsf{m}_u}}}^{\text{exiting } \mathsf{m}_u}
    \cdot \overbracket{\frac{q_{\mathsf{m}_v}}{q}}^{\text{entering } \mathsf{m}_v}
    \cdot \overbracket{\frac{p_v}{p_{\mathsf{m}_v}}}^{\text{visiting node } v} \Bigr) \Biggr)
    + \left( \sum_{u \in \mathsf{m}} p_u \sum_{v \in \mathsf{m}} t_{uv} \log_2 \frac{p_v}{p_{\mathsf{m}_v}} \right).
\end{align*}
We realise that, for each module $\mathsf{m}$, we sum over all nodes $u \not\in \mathsf{m}$ and all nodes $u \in \mathsf{m}$, and, depending on whether they are a member of $\mathsf{m}$, calculate the cost for transitioning to $v \in \mathsf{m}$ differently.
We rewrite these two cases using the Kronecker delta $\delta$, summing over all nodes $u$,
\begin{align*}
    - \sum_{\mathsf{m} \in \mathsf{M}} \sum_u p_u \sum_{v \in \mathsf{m}} t_{uv} \left[ \left(1-\delta_{\mathsf{m}_u, \mathsf{m}_v} \right) \log_2 \left( \frac{q_{\mathsf{m}_u}}{p_{\mathsf{m}_u}} \cdot \frac{q_{\mathsf{m}_v}}{q} \cdot \frac{p_v}{p_{\mathsf{m}_v}} \right)
    + \delta_{\mathsf{m}_u, \mathsf{m}_v} \log_2 \frac{p_v}{p_{\mathsf{m}_v}}
    \right].
\end{align*}
Finally, instead of summing over all modules and all nodes in each module, we sum over all nodes directly and can calculate the codelength for partition $\mathsf{M}$ as
\begin{align*}
    L\left(\mathsf{M}\right) & = - \sum_u p_u \sum_v t_{uv} \left[ \left(1-\delta_{\mathsf{m}_u, \mathsf{m}_v} \right) \log_2 \left( \frac{q_{\mathsf{m}_u}}{p_{\mathsf{m}_u}} \cdot \frac{q_{\mathsf{m}_v}}{q} \cdot \frac{p_v}{p_{\mathsf{m}_v}} \right)
    + \delta_{\mathsf{m}_u, \mathsf{m}_v} \log_2 \frac{p_v}{p_{\mathsf{m}_v}}
    \right] \\
    & = - \sum_u p_u \sum_v t_{uv} \log_2 \left[ \left(1-\delta_{\mathsf{m}_u, \mathsf{m}_v} \right) \left( \frac{q_{\mathsf{m}_u}}{p_{\mathsf{m}_u}} \cdot \frac{q_{\mathsf{m}_v}}{q} \cdot \frac{p_v}{p_{\mathsf{m}_v}} \right)
    + \delta_{\mathsf{m}_u, \mathsf{m}_v} \frac{p_v}{p_{\mathsf{m}_v}}
    \right].
\end{align*}

The part inside the square brackets is known as map equation similarity, or \emph{mapsim} for short, an information-theoretic measure for node similarity~\cite{pmlr-v198-blocker22a}.
That is, we can calculate the codelength for a partition $\mathsf{M}$ using $\operatorname{mapsim}$, where $\log_2 \operatorname{mapsim}\left(\mathsf{M}, u, v\right)$ quantifies how many bits are required to describe a random-walker-transition from node $u$ to $v$, given partition $\mathsf{M}$,
\begin{align*}
    L\left(\mathsf{M}\right) = - \sum_u p_u \sum_v t_{uv} \log_2 \operatorname{mapsim}\left(\mathsf{M}, u, v\right).
\end{align*}

\end{document}